\documentclass{article}

\usepackage{PRIMEarxiv}

\usepackage[utf8]{inputenc} 
\usepackage[T1]{fontenc}    
\usepackage{hyperref}       
\usepackage{url}            
\usepackage{booktabs}       
\usepackage{amsfonts}       
\usepackage{nicefrac}       
\usepackage{microtype}      
\usepackage{lipsum}
\usepackage{fancyhdr}       
\usepackage{graphicx}       
\graphicspath{{media/}}     
\usepackage{booktabs} 
\usepackage{tabularx} 
\usepackage{float}
\usepackage{niceframe} 

\title{Building a Method: Challenges and Lessons from Developing an Impact Evaluation Framework for Public Investment in Open Source Software
}

\author{
  Domenech Burin, Laia \\
  Data Scientist \\
  Sovereign Tech Agency \\
  Berlin\\
  \texttt{laia@sovereign.tech} \\
}

\begin{document}
\maketitle

\begin{abstract}
Modern societies and institutions depend on an  invisible layer of open digital infrastructure: foundational, openly accessible digital technologies that support a broad range of social and economic activities. Open source software is a core component of this infrastructure, underpinning everyday technologies and critical systems in various critical sectors. As open source software becomes increasingly recognised, funded, and governed as open digital infrastructure, policymakers face growing demands for accountability, transparency, and credible evidence of impact.

However, methods for evaluating public investment in open source software remain underdeveloped, with no widely accepted methodological standard. This paper describes the Sovereign Tech Agency’s transition from qualitative reporting and funding-recipient self-assessment towards a mixed-methods approach that incorporates quantitative, goal-oriented evaluation to inform funding decisions. Drawing on the development of this framework, it examines its initial design, the practical and structural challenges encountered, and the responses adopted.

The work identifies three central challenges: evaluation indicators must reflect heterogeneous project goals and contexts; quantitative assessment relies on fragmented and fragile data infrastructure; and observed changes before, during, and after investment do not establish causal attribution without additional research designs. The paper argues that evaluating investment in open source software  is not only a methodological task but also a governance practice, dependent on sustainable open research infrastructure, community knowledge, and ongoing methodological refinement.
\end{abstract}


\section{Introduction}

Open source software (OSS) is software whose source code is made available under a licence that permits anyone  to inspect, use, modify, and redistribute it.  It is often developed and maintained collaboratively and in a decentralized manner, and is typically available at no cost. OSS underpins much of the digital infrastructure on which contemporary society and the economy depend. Over 96\% of codebases contain OSS (\cite{synopsys_open_source_security_and_risk_analysis_deep_2023}), and roughly 70-90\% of any software stack consists of OSS (\cite{open_source_security_foundation_open_2022}). 

While used by many, projects are often maintained by small groups of people or, in some cases, individuals. Despite its criticality and widespread use, OSS maintenance work remains inadequately compensated. Those in the open source community have increasingly been expressing concerns \cite{salkever_open_2023} around widespread commercial freeriding on the labour of volunteers \cite{geiger_labor_2021}, security vulnerabilities and maintainer burnout \cite{asparouhova-working}, both of these fueled by the increasing appearance of AI reporting \cite{learner_who_2026}.

This prevalence of OSS across modern codebases, and its presence in various systems of critical infrastructure such as finance and energy \cite{nagle_strengthening_2022} has given it the recognition as open digital infrastructure (ODI). ODI refers to foundational and openly accessible digital technologies that make crucial contributions to society, serving as input resources for a broad range of downstream productive activities \cite{luhrsen_organizing_2025}. As the infrastructural role of OSS has gained recognition, its sustainability has become an increasingly important public policy concern. In response, various governments have adopted measures to address this issue, primarily through direct investment in OSS maintenance. Recent European examples include the Next Generation Internet (NGI), funded by the European Comission, the Sovereign Tech Agency in Germany and the European Sovereign Tech Fund pilot implemented through the EDIC Digital Commons.

With the growth of budgetary expenditure in OSS, the question of measurement becomes increasingly important: how can public institutions determine that public investment strengthens those open source components? Evaluating the results of investment can provide insights into what kind of instruments work, which ones don’t, and how changes in the well-being of the projects can be attributed to the different interventions. Establishing causality in those changes, however, requires complex research designs. As such, impact evaluation for OSS funding is an ongoing research area, with no widely accepted methodological standard.

Despite this methodological uncertainty, public funders face growing demands for quantitative reporting to keep stakeholders and taxpayers informed about the progress and outcomes of public investment. Agencies investing in OSS must therefore develop evaluation methods that meet these accountability demands while remaining explicit about what the evidence can—and cannot—establish.

This paper examines the development of an evaluation framework developed by the Sovereign Tech Agency rather than presenting a completed evaluation of funding effects. Drawing on the process of developing and testing the framework, its contribution is to identify methodological and infrastructural challenges, and explain how to overcome them in the framework’s design. The work describes the development of a context-sensitive, mixed-methods approach to evaluating public investment in open source software. 

\section{Review of Practice \& Literature Review}
\label{sec:rop_litr}

\subsection{Literature Review}

Public sector management has increasingly emphasised linking public expenditure to explicit objectives and measurable results. Often associated with the “results agenda”, this approach refers to the notion that public interventions should be evaluated based on predetermined results through the use of quantified and aggregated indicators \cite{vallejo_capacity_2016}. This concept is also tied to the idea of evidence-based policy: it is possible to use rigorous information to support different policy decisions under the premise of showing “governamental effectiveness”  \cite{davies_viewpoint_1999} through successful results and evaluation to improve policies \cite{sanderson-evaluation}.

In Germany, this broader shift is reflected in the move towards \textit{ziel- und wirkungsorientierte Haushaltsführung} (zwoH), which seeks to connect budgetary decisions more systematically to the goals, outcomes, and impacts of public expenditure. This was emphasised in a joint declaration by the Federal Government through the Ministry of Finance’s 12th Spending Review \cite{bundesministerium_der_finanzen_gemeinsame_2025}, alongside the publication of a framework outlining the general concept and objectives of an ideal zwoH. The commitment endorsed by the three parties of the governing coalition positions this approach as an explicit federal policy priority \cite{bundesministerium_der_finanzen_ziel-_2025}. For public institutions investing in OSS, this creates a corresponding need for methods that can generate credible evidence about the results of their interventions.

In this goal-oriented paradigm, quantitative methods take precedence over qualitative approaches, despite their inherent limitations in capturing the depth and context of social phenomena. Demonstrating the impact of public programmes through quantitative indicators is something that can be achieved through monitoring and impact evaluation. Monitoring a programme involves tracking indicators of progress as a basis on which to evaluate outputs and outcomes \cite{khandker-handbook}. Outputs refer to the supply-side deliverables, such as the events, products, or services that result from an investment intervention. Outcomes refer to the likely or achieved short-term and medium-term effects of an intervention’s outputs, such as increased security or stronger maintenance \cite{world_bank_evaluation_2020}. 

On the other hand, impact evaluation studies whether the changes in a targeted population (in our case, open source projects) are due to the intervention and not other factors. Monitoring looks at the changes of different metrics, impact evaluation intends to demonstrate that changes in targets are due to the specific intervention undertaken \cite{khandker-handbook}. The latter is achieved through causal identification, which is anchored in the potential outcomes framework (\cite{rubin_causal_2005}). This framework proposes comparing what would happen to a unit of analysis if it was treated and if it wasn’t treated. However, the core problem is that there is no scenario where both cases exist. Therefore, evaluation programmes must adopt experimental or semi-experimental approaches to draw conclusions on the effects of funding.

Applying these approaches to OSS presents various challenges. Open source projects are heterogeneous sociotechnical systems whose technical characteristics, governance structures and communities vary substantially. Additionally, the types of public intervention also vary in objectives and format. As a result, research cautions against relying solely on quantitative methods for impact assessment. Because open source projects are heterogeneous sociotechnical systems, qualitative methods can explain how investment interacts with a project’s particular circumstances. They can reveal unexpected effects, illustrate complex interactions between funding and project dynamics, and support the interpretation of observed changes. Osborne et al. \cite{osborne-toolkit} highlight the complexity of assessing impact across internal and external dimensions, as well as economic, social, and technological domains. They favour mixed-methods approaches that combine scalable measurement with contextual depth.

Previous evaluations of Sovereign Tech Agency programmes have relied on qualitative assessments of reported outcomes and interviews with participating maintainers. To date, its benchmark evaluation reports are the Sovereign Tech Fund Pilot Evaluation and the Fellowship Evaluation report. In the ZEW and Agora Digital report The State of Goal- and Impact-Orientation in Digital Policy, the agency scores 3.8 out of 5 for adopting impact-oriented policy. The main gaps identified are the lack of quantitative assessment in impact reporting and limited data openness \cite{heinemann_digital_2026}.

\subsection{Review of Practice: Sovereign Tech Agency}
\label{sec:rop}

The Sovereign Tech Fund (STF) was established in 2022 by the German Federal Ministry for Economic Affairs and Climate Action (BMWK) and hosted by SPRIND GmbH, the Federal Agency for Disruptive Innovation, to strengthen the resilience, security, and sustainability of critical open source digital infrastructure. The STF provides OSS projects with contracts for maintenance and development work starting from €50,000, with no upper funding limit, and supports project durations ranging from 6 to 24 months. In November 2024, the STF was reorganised under a newly formed umbrella entity: the Sovereign Tech Agency (STA). This transition allowed the STA to expand its scope beyond the STF’s original focus, establishing a broader portfolio of programmes supporting open digital infrastructure

Since its inception, the STF has provided €39 million in funding to more than 110 critical open source projects. The contracted work varies substantially across projects: from the implementation of new features to the components, producing security guidelines or compliance tools, allocating time to reducing existing backlog through bug clean-ups and doing community work. These different kinds of contracts also produce very different outcomes: such as increasing security within a project, strengthening the communities or modernising the software.

\subsection{Review of Practice: Policy and Research at the Sovereign Tech Agency}

Within the Sovereign Tech Agency, the Policy \& Research team develops evidence to support both institutional accountability and learning about the effects of public investment in open digital infrastructure. This works responds to the broader shift towards goal- and impact-oriented public spending described above, while also addressing a practical question for the Agency: how can evidence about contracted work on projects be used to understand whether interventions are achieving their intended outcomes?

The team combines externally commissioned research is used to explore questions that require additional methodological expertise, analysis, or data beyond the Agency's own capacity. In parallel, the team is developing an internal mixed-methods evaluation framework, initially focused on the Sovereign Tech Fund, that can be integrated more systematically into the funding process. 

The framework is intended to support three related goals. First, it should strengthen the Agency's accountability and demonstrate its efficacy by providing evidence about the outcomes associated with public investment in OSS. Second, it should enable structured comparision across projects where their goals, contracted work types, and available data make such comparison meaningful. Third, it should eventually support learning within the Agency by identifying whether expected changes occur, how these changes differ across types of intervention, and where further investigation, qualitative or causal, is required. 

Developing this framework requires balancing standardisation with the heterogeneity of open source projects. Rather than applying a fixed set of indicators across the portfolio of projects, the approach starts from the objectives of each funded intervention and seeks to translate them into appropriate evaluation questions and observable metrics. The framework presented in this paper emerges from that work and draws on the author’s experience as a practitioner involved in its development.

\section{Building the Methods}

\subsection{Design Challenges}

Evaluating public investment in OSS presents several interrelated methodological challenges, beginning with the sociotechnical complexity of individual projects. These challenges become more pronounced when assessing effects across entire ecosystems. Addressing this complexity through qualitative approaches is feasible, and allows for researchers and public agencies to better understand the results of the investments on a case-by-case basis. However, qualitative approaches do not allow for standarisation and scalability of results. This raises the question of how to develop quantitative approaches to evaluate the impact of investment in open source, and what considerations researchers should take into account when developing them. This section presents the main challenges that arised when doing this research in practice as a public state agency: unit definition, data infrastructure, impact definition and inference. 

The choice of unit of analysis is the starting point: it guides where to look for data, how to define impact, and how to identify comparable units. The term “open source project” encompasses various sociotechnical dimensions of the software, including its community, repositories, organisational structures, and technical dependencies. These dimensions do not necessarily align neatly with one another. Operationalising the term therefore requires distinguishing between them and specifying which level of the project is being assessed. This definition shapes what can be observed and the conclusions that can be drawn.

Data is the main input to produce quantitative research. In this aspect, data from open source projects also presents challenges and limitations. The main challenge is the fragmentation across different sources: development, documentation and contribution activity is found in repositories that can be hosted in different version control platforms. While GitHub remains the biggest and most popular provider to host git-based code repositories \cite{wikipedia_github_2026}, open source projects might also rely on alternatives like GitLab or Bitbucket. Additionally, vulnerabilities or bugs can be reported in private bug managers from the different projects before they are disclosed publicly in the CVE catalog. As a result, the data collection pipeline must take into account multiple sources. Or, if it focuses in one of these sources, address the limitations of analysis.

In addition, there is information related to the project’s community and governance structure that is not publicly disclosed or expressed in the aforementioned sources. Mailing lists, internal discussions on instant messaging platforms, are sources that can capture those dynamics more comprehensively. However, this appears in the format of text data, and can only be analysed qualitatively or through Natural Language Processing (NLP) approaches. Those approaches therefore require complementary qualitative or unstructured-data approaches and are not within the scope of this paper, as the focus is in structured data analysis.

Summarising this data into metrics that are relevant to identify impact of the investment is the next stage of research. The literature on impact measurement in open source, together with the nature of the Sovereign Tech Fund as an instrument, suggests that no single set of metrics is suitable for all projects. Interpreting the metrics in the context of projects is an integral part of the process. Firstly, because changes in individual metrics can be ambiguous. For example: an increase in disclosed CVEs may reflect improved detection and disclosure rather than deteriorating security, and fewer commits may indicate a reduced maintenance burden rather than declining activity. 

Secondly, the type of work done with the Fund's investement varies across the kind of maintenance needed by the project. This can encompass software development, security audits, bug fixes, and community-based work. Subject-matter expertise is therefore needed to understand what each work package entails and select software health indicators that meaningfully capture its intended outcomes.

Establishing that the changes produced in the metrics are a result of the investment intervention is a challenge of causal attribution. The potential outcomes framework, introduced in the literature review, underpins causal assessments by asking: what would have happened without the investment? Answering this requires constructing potential counterfactuals, either through experimental or quasi-experimental designs. However, randomised experiments are unlikely for these policy instruments because funding decisions involve predefined selection based on institutional knowledge and budget constraints. Attribution is further complicated by overlapping funding sources, external changes (such as shifts in governance or new technological developments) and dependencies or spillover effects across projects.

The following section describes how the different challenges were addressed and the outcomes of it.

\subsection{Methodological Framework}

The framework was developed iteratively through project-level case studies combining qualitative information about funded work with quantitative analysis of repository and ecosystem data. Four projects from the Fund’s pilot round were selected for the initial studies. The sample was chosen on the criteria of time and work packages. For the first one, enough time had passed since the investment to observe medium- to long-term changes. For the second aspect, they differed substantially in their sociotechnical structures and intervention goals, which allowed to capture the heterogeneity of projects and work done with the Fund's investment. After this initial round, a fifth project, the OpenJS Foundation, was added because its work package focused on community-related activities that led to security and technical developments. This broadened the analysis to reflect the diverse range of maintenance work undertaken in open source and funded by the Agency.   

\begin{itemize}

\item \textbf{Python Package Index (PyPI):} Python is one of the most widely used programming languages for developing websites, software, and machine learning and AI models in today's development landscape (\cite{python_software_foundation_python_2024}). Package management in Python is handled through the pip ecosystem, which hosts its packages on PyPI. With over 1,046,786 users and 794,245 projects, it stands out as one of the primary package repositories for Python (\cite{python_package_index_pypi_nodate}). The first round of investment during the pilot of the Sovereign Tech Fund focused primarily on implementing new features to strengthen the overall ecosystem. The latter investment round, dating to end of 2025, targeted work in systematic security analysis and experimental development.

\item \textbf{curl:} curl is an open source tool used in command lines or scripts to transfer data. The library behind it, libcurl, is one of the world’s most used libraries. It is widely used from countless software applications to cars, television sets, routers, printers, audio equipment, mobile phones, tablets, medical devices, and more. It has over twenty billion installations (\cite{curl_curl_nodate}). The funding of the STF targeted maintenance via addressing technical backlog: resolving 122 known bugs through systematic classification and fixes, while also introducing new options to support newer versions of the HTTP protocol.

\item \textbf{Fortran:} Fortran is a widely used programming language that has been designed from the ground up for computationally intensive applications in science and engineering. Today, it is the standard in domains that adopted computation early–science and engineering. These include numerical weather and ocean prediction, computational fluid dynamics, applied math, statistics, and finance. It is particularly well-established in scientific communities due to its reliability (\cite{fortranlang_fortran_nodate}). STF investment focused on modernizing the Fortran ecosystem through improvements to the compiler and the package manager, ensuring continued usability and relevance for contemporary scientific computing.

\item \textbf{RubyGems:} RubyGems and Bundler provide library and software package management for the Ruby programming language, included with every copy of Ruby in the language standard library. They are indispensable tool for using Ruby, and many applications rely on their regular maintenance and improvement, as well as their ongoing usability \cite{rubygems_bundler_nodate}.
RubyGems.org hosts publicly available software packages (called “gems”) which enables developers to quickly and securely integrate more functionalities into their projects, while Bundler downloads and installs the exact gems and versions needed for a given project. The STF funding was used to strengthen community maintenance, improving the reliability, security and scalability of RubyGems.org and improve maintainer quality of life by reducing repetitive work and risk (via automation, clearer tools, and safer defaults), giving maintainers and contributors better visibility, workflows, and support systems to collaborate effectively.

\item \textbf{OpenJS Foundation:} The OpenJS Foundation supports foundational JavaScript projects and related web technologies, including Node.js, jQuery, and webpack. Its first Sovereign Tech Fund contract focused on strengthening the security and reliability of this ecosystem while reducing the infrastructure and maintenance burden on project maintainers. The commissioned work combined shared build, test, and deployment infrastructure with security engineering support, including adapting OpenSSF and OWASP best practices to JavaScript, developing security guidance and training, and improving secure release processes and CVE management. It also included security audits and an infrastructure inventory to support the responsible transition of inactive projects to Emeritus status. The contract thus linked community coordination and shared maintenance capacity with technical improvements across multiple projects.

\end{itemize}

In response to the unit-specification challenge described, the initial analysis treated the repositories where the commissioned work was carried out as the primary units of analysis. The material used to inform the quantitative metrics were qualitative sources. Those were the contracts, which outlined the different work packages; the reports produced by the projects after completion; and interviews with maintainers. These defined the construction and interpretation of the quantitative metrics.

These work packages were translated into metrics using the Goal-Question-Metric (GQM) approach \cite{marciniak_goal_2002}. GQM is a structured, top-down framework for defining and evaluating meaningful metrics. It involves identifying a clear objective for the process, formulating questions to assess whether that objective is being achieved, and defining metrics to answer those questions. This was particularly useful given the heterogeneity of the commissioned work: rather than imposing a common set of portfolio-wide KPIs, the framework begins with the purpose of each intervention and derives project-specific indicators from it. Therefore, following this framework, the work packages were analyzed across these dimensions and translated into metrics.

\begin{table}[htbp]
\centering
    \small
    \caption{Project-specific metrics derived using the
    Goal--Question--Metric (GQM) approach.}
    \label{tab:gqm-project-metrics}
    \begin{tabularx}{\linewidth}{@{}
        >{\raggedright\arraybackslash}p{0.13\linewidth}
        >{\raggedright\arraybackslash}X
        >{\raggedright\arraybackslash}X
        >{\raggedright\arraybackslash}X
        @{}}
        \toprule
        \textbf{Project}
        & \textbf{Goal}
        & \textbf{Evaluation question}
        & \textbf{Example metrics} \\
        \midrule
        \textbf{PyPI}
        & Improve ecosystem tooling and security.
        & Did the funded work accelerate the adoption of improved
          tooling across the PyPI ecosystem?
        & Commits, change requests, OpenSSF security signals \\
        \addlinespace

        \textbf{Fortran}
        & Improve the usability of the Fortran ecosystem.
        & Did the funded work make the package registry more
          operationally useful?
        & Median issue-resolution time, time to first response \\
        \addlinespace

        \textbf{RubyGems}
        & Improve maintainer quality of life.
        & Did the funded work improve maintainer experience and
          community engagement?
        & Median issue-resolution time, new issues, median
          change-request duration \\
        \addlinespace

        \textbf{curl}
        & Reduce the maintenance backlog.
        & Did the funded work help resolve the accumulated bugs?
        & Closed issues, median issue-resolution time \\
        \addlinespace

        \textbf{OpenJS Foundation}
        & Improve security in the JavaScript ecosystem.
        & Did the funded work strengthen security guidance and tooling?
        & Disclosed CVEs, commits in security working groups and
          compliance-tooling repositories \\
        \bottomrule
    \end{tabularx}
\end{table}

GQM resulted the best approach to define concrete questions related to the work packages. The elaboration of metrics, however, was not only a product of the internal research work at the Agency. They were selected based in expert knowledge from open source communities. These communities have internalized knowledge on their dynamics, and which are the indicators that inform on a project’s health. Therefore, to achieve construct validity, the research team relied on the work developed by the CHAOSS Community to define metrics. CHAOSS is a Linux Foundation project that creates metrics, metrics models, and software to better understand the health of open source communities globally \cite{chaoss-chaoss}. CHAOSS’ metrics models were the main input to map work packages into different quantitative indicators, using various indicators per work package to identify health across various dimensions. 

This community-rooted approach was complemented through a second stage of interviews and conversations with members of the CHAOSS community and the projects themselves. The projects were informed that they would be included as evaluation case studies, and the proposed metrics are discussed with them on an ongoing basis to gather their insights into the nuances of each metric and their perspectives on what constitutes a successful impact. This exercise displayed an additional layer of complexity when measuring impact of investment: the distinction between outputs and outcomes of a project. The power dynamic between the Agency as a commissioner and the projects contracted requires making explicit that the goal is to evaluate the efficacy of the policy instrument, not the projects themselves. Otherwise, the question “what metrics are useful to see the long-term health of your project, related to the working package?” might be understood as “what are output results we can see from the work you have done?”. This fundamentally alters the goal of impact assessment.

Data collection also relied on research collaborations and tools developed by and for the open source community. The primary data source was \href{https://ecosyste.ms/}{ecosyste.ms}  \cite{noauthor_ecosystems_nodate}, the largest dataset on open source production. Collaboration and dialogue with its core maintainer proved especially valuable and enabled the development of new tools. Security-related indicators, for example, are important but difficult to capture over time. OpenSSF’s Security Scorecard Checks \cite{noauthor_ossfscorecard_2026} provide benchmarks for assessing whether a repository follows good security practices, but historical time-series data for these checks are unavailable. To address this gap, the \href{https://github.com/git-pkgs/signals}{signals} tool was developed to trace repository-based security signals across Git history. It records when evidence of these practices appears, changes, or disappears, together with the corresponding commit and file path. Linking these signals to timestamped commits makes it possible to track observable security practices over time and compare them before and after investment. 

These data sources and tools form research infrastructure, which shapes what can be observed in the open source ecosystem. However, this infrastructure shares many of the vulnerabilities of the projects it enables researchers to study. Dependence on a single maintainer, compounded by inadequate financial support, threatens its sustainability and, in turn, the continuity of the research it supports.

Combining these data sources and metric models, the team developed a quantitative descriptive approach to assess the effects of the Sovereign Tech Fund's investments. This approach applied on the case studies elaborates descriptive time series visualizations of the defined metrics, highlighting changes in them before, during and after investment. Combined with qualitative contextualisation and analysis, these descriptives present the agency’s goal-oriented results. 

The described approach allows to describe changes around investment, and links the Sovereign Tech Fund’s policy objectives to projects’ work packages, intended outcomes, and observed changes. It illustrates how a funder can move from policy objectives to defensible evidence while intentionally remaining explicit about the limits of inference and the conclusions that can be drawn. However, it does not resolve the challenge of causal inference. Future work within the team aims to address this. Initial research by Domenech Burin \cite{burin-measuring} applies Generalized Synthetic Control methods to these case studies to estimate the programme’s causal effects.

\section{Conclusion}

This working paper presented the Sovereign Tech Agency’s approach to build an impact evaluation framework for public investment in open source software. No generalised method exists so far in the research literature or policy practice, and the experience shows that a one-size-fits-all solution for this is unlikely to be found. Due to the heterogeneity of open source projects and their maintenance, different shapes and formats of investment instruments and the fast-changing nature of technology, fueled by developments in artificial intelligence, a static evaluation framework can become obsolete quickly. 

The main challenges encountered during framework development were the heterogeneity of funded activities and their intended outcomes, the fragmentation of data sources, the translation of broad work packages into meaningful metrics, and the difficulty of attributing observed changes to investment. In response, a context-sensitive quantitative approach within an iterative mixed-methods framework was developed. This framework combines goal-oriented metric selection, engagement with ecosystem stakeholders, and collaboration on data infrastructure. This experience highlighted the importance of developing methods that are adaptable across contexts, transparent about assumptions and limitations, and co-developed with ecosystem stakeholders.

\textbf{Adaptable across contexts:} Methods should reflect the context of both the commissioning instrument and the open source projects. Clearly identifying the instrument’s goals, the type of maintenance to be undertaken, and each project’s context is crucial for defining what impact should look like and translating it into metrics. Mixed-methods approaches provide a strong foundation for gathering qualitative contextual information, while GQM can translate these insights into concrete metrics, balancing context with measurability.

\textbf{Transparency in assumptions and limitations:} Any scientific attempt to capture a phenomena requires abstraction from reality. The same applies to developing this methodological framework: even mixed-methods approaches rely on assumptions about why the chosen metrics capture the relevant dimensions. Decisions about the unit of analysis and data sources introduce bias into what of the open source project is being measured. Making these choices and assumptions explicit is therefore essential to ensure transparency for stakeholders and in the use of public funds. 

\textbf{Co-development with ecosystem stakeholders:} Research on open digital infrastructure policy should involve researchers and communities from the open source ecosystem. These communities hold knowledge grounded in their practices and history, and preserving it is essential to the impact evaluation framework. Research collaborations, expert review of the methods, and interviews with the projects being evaluated are integral to developing the framework. These engagements must account for power imbalances between funders and funding recipients, particularly during interviews, when recipients may feel constrained in offering critical feedback or discussing challenges openly.

The Sovereign Tech Agency’s experience suggests that sustaining open source as digital infrastructure requires more than investing in individual projects: it also requires shared, legible, and adaptable systems for understanding the outcomes of those investments. The outlook for the team is to continue building on the developed framework, integrating it into the programmatic work. The goal is to include metrics from the moment a project enters the Soveregin Tech Fund pipeline, through the application process. 

A complementary research priority is to strengthen causal assessment and investigate how evaluation can extend beyond selected case studies to programme-level effects. The descriptive approach presented here provides a foundation for this work, but establishing attribution will require additional research designs and explicit assumptions. 

Public institutions investing in open digital infrastructure should not assess impact through a static set of quantitative indicators alone. Instead, they should approach impact evaluation as a shared governance practice grounded in openness and collaboration with the communities that develop and maintain this infrastructure. Such an approach can lead to methods that balance accountability for the use of public funds with a deeper understanding of the sociotechnical dynamics that shape investment outcomes.

\bibliographystyle{unsrt}  
\bibliography{references}

\end{document}